\documentclass[11pt]{article}

\usepackage{indentfirst}
\usepackage{graphicx}
\usepackage{longtable}
\usepackage{supertabular}
\usepackage{amsmath}
\usepackage{amssymb}
\usepackage{color}%for text color
\usepackage{amstext}
\usepackage{cases}
\usepackage{float}%insert a picture to allocated place
\usepackage{threeparttable}
\usepackage{booktabs}%In order to use \toprule,\midrule,\bottomrule
\usepackage[margin=.8in]{geometry}
\usepackage{setspace}
\usepackage{multirow}
\usepackage{array}
\usepackage{mathrsfs}

\begin{document}

\title{{\Large\bf {Describing Sen's Transitivity Condition by Use of Inequality and Equation}} }

\author{Fujun Hou \thanks{Tel.: +86 10
6891 8960; Fax: +86 10 6891 2483; email: houfj@bit.edu.cn.}\\
%EndAName
School of Management and Economics\\
Beijing Institute of Technology\\
Beijing, China, 100081\\
}
\date{\today}
\maketitle

\begin{abstract}
In social choice theory, Sen's value restriction condition is a sufficiency condition restricted to individuals' ordinal preferences so as to obtain a transitive social preference under the majority decision rule. In this article, Sen's transitivity condition is described by use of inequality and equation. First, for a triple of alternatives, an individual's preference is represented by a preference map, whose entries are sets containing the ranking position or positions derived from the individual's preference over that triple of those alternatives. Second, by using the union operation of sets and the cardinality concept, Sen's transitivity condition is described by inequalities. Finally, by using the membership function of sets, Sen's transitivity condition is further described by equations.

{\em Keywords}: social choice theory, social preference transitivity, Sen's value restriction condition, preference map
\end{abstract}

\setlength{\unitlength}{1mm}

\section{Introduction}

If the number of alternatives is more than two, Arrow proved that, it is impossible for the majority decision rule to yield a transitive social ordering under the requirement that Arrow's conditions 2-5 are satisfied (Arrow, 1950, 1951). This famous "Impossibility Theorem" has a profound impact on the development of modern social choice theory. To make the majority decision rule yield a transitive social preference, researchers attempted to impose some constraints on the individuals' preferences patterns. Thus there have appeared many transitivity conditions among which Sen's value restriction condition (Sen, 1966) is a representable one since it covers several well-known sufficiency conditions that have preceded it such as the single-peaked preferences pattern (Arrow, 1951; Black, 1958), the single-caved preferences pattern (Inada, 1964), the dichotomous preferences pattern (Inada, 1964), the two-group separated preferences pattern (Inada, 1964), and the Latin-square-lessness preferences pattern (Ward, 1965).

Sen's condition does its work by requiring that in any triple of alternatives at least one alternative not have the value 'best', 'worst', or 'medium'. Therefore, Sen's condition is a qualitative one. This article is concerned with how to describe Sen's condition in a quantitative manner. In particular, we will characterize it by use of inequality and equation. Quantitative description can be a great help for examining Sen's condition with a computer.

\section{Preference map and membership preference map}

When it comes to the social preference transitivity, Arrow and Sen (among others) are concerned primarily with a scenario that the individuals' preferences over a finite alternative set are weak orderings (Arrow, 1951; Sen, 1970). A weak ordering on a finite set can be and will be here represented by a preference map. This is crucial for the discussion in this article.

The preference map representation method assumes that the alternatives in a tie are tied together occupying common positions, and these positions are consecutive positive integers. Denote by $A=\{x_1,x_2,\ldots,x_m\}$ the alternative set with $1<m<+\infty$ and by
$N=\{1,2,\ldots,n\}$ the individual group with $1<n<+\infty$. The preference map is defined as follows (Hou, 2015a, 2015b; Hou \& Triantaphyllou, 2019).

\textbf{Definition 1} A sequence $(PM_i)_{m\times 1}$ is called a \textit{preference map} (PM) with respect to a weak order relation $\preceq$ on the alternative set $A=\{x_1,x_2,\ldots,x_m\}$ such that
$$PM_i=\{|\xi_i|+1,|\xi_i|+2,\ldots,|\xi_i|+|\eta_i|\},\eqno(1)$$
where the notation $|\centerdot|$ represents the cardinality of a set;
$\xi_i$ is the \textit{predominance set} of alternative $x_i$, i.e., $\xi_i=\{x_k\mid x_k\in A,x_i\prec x_k\}$; and $\eta_i$ is the \textit{indifference set} of alternative $x_i$, i.e., $\eta_i=\{x_k\mid x_k\in A,x_i\sim x_k\}$.

For illustrative purpose, we consider an example.

\textbf{Example 1} Suppose that three voters provide their preferences over an alternative triple $\{x,y,z\}$ as follows.
$$\begin{array}{ll}
 \text{Voter 1}: & x\succ y\succ z,\\
 \text{Voter 2}: & x\sim y\succ z,\\
 \text{Voter 3}: & x\sim y\sim z.
\end{array}
$$
According to Def.1, the voters' preferences can be represented by the following
preference maps
$$
\begin{array}{c}{\hspace{0.3cm}\begin{array}{cccc}\text{Voter 1}:
&\hspace{0.2cm}\text{Voter 2}: &\hspace{0.4cm}\text{Voter 3}:
\end{array}}\\
{\begin{array}{cc}
{\begin{array}{c} x\\
y\\
z
\end{array}}&
{ \left[\begin{array}{c}
\{1\}\\
\{2\}\\
\{3\}\end{array}\right],
\left[\begin{array}{c}\{1,2\}\\
\{1,2\}\\
\{3\}\end{array}\right],
\left[\begin{array}{c}\{1,2,3\}\\
\{1,2,3\}\\
\{1,2,3\}\end{array}\right],
}
\end{array}}
\end{array}
$$
from which one can see the alternatives' ranking positions corresponding to the voters' preferences. For instance, voter 1 would like to rank alternative $x$ at position 1, while voter 2 would like to rank alternative $x$ at position 1 or position 2, and positions 1-3 are all agreed by voter 3 with regard to alternative $x$'s rank order. According to Sen (1966), by the way, voter 3's preference in Example 1 is a sort of \textit{unconcerned} preference over the triple $\{x,y,z\}$. A unconcerned individual is a person who is indifferent between all the alternatives (Sen, 1966), and accordingly, the preference of a unconcerned person is called unconcerned preference. In contrast, a \textit{concerned} individual is a person who is not indifferent between all the alternatives (Sen, 1966).

Because the preference map entries are sets rather than numbers, a preference map will have a one-to-one membership correspondence which is none other than a 0-1 matrix as defined below.

\textbf{Definition 2} For a preference map $(PM_i)_{m\times 1}$, its \textit{membership preference map} (mPM) is a $m\times m$ 0-1 matrix $(mPM_{i,j})_{m\times m}$ whose elements are defined as
$$
{mPM_{i,j}}=\begin{cases}
1, & \text{if $j\in PM_i$},\\
0, & \text{otherwise}.
\end{cases}
\eqno(2)$$

When the previous data are used, then the corresponding mPMs of the PMs in Example 1 are as follows:
$$
\begin{array}{c}{\hspace{0.3cm}\begin{array}{cccc}\text{Voter 1}:
&\hspace{0.2cm}\text{Voter 2}: &\hspace{0.4cm}\text{Voter 3}:
\end{array}}\\
\begin{array}{c}{\hspace{0.8cm}\begin{array}{cccc}\{1\}\hspace{0.08cm}\{2\}\hspace{0.08cm}\{3\}
&\hspace{0.1cm}\{1\}\hspace{0.08cm}\{2\}\hspace{0.08cm}\{3\} &\hspace{0.1cm}\{1\}\hspace{0.08cm}\{2\}\hspace{0.08cm}\{3\}
\end{array}}\\
{\begin{array}{cc}
{\begin{array}{c} x\\
y\\
z
\end{array}}&
{ \left[\begin{array}{ccc}
1 &  0 &  0 \\
0 &  1 &  0 \\
0 &  0 &  1
\end{array}\right],
\left[\begin{array}{ccc}
1 &  1 &  0 \\
1 &  1 &  0 \\
0 &  0 &  1
\end{array}\right],
\left[\begin{array}{ccc}
1 &  1 &  1 \\
1 &  1 &  1 \\
1 &  1 &  1
\end{array}\right],
}
\end{array}}
\end{array}
\end{array}
$$
where the left column indicates the alternatives, the first row indicates the voters, the second row indicates the ranking positions, and a 0-1 value indicates whether a certain alternative can or cannot be ranked at a certain position by a certain voter.

\section{Sen's condition in inequalities}

We start in this section by introducing Sen's transitivity condition and some underlying concepts. Thereafter, we describe Sen's transitivity condition in inequalities and illustrate an example.

\textbf{Theorem 1 (Sen's transitivity condition)} \textit{The
method of majority decision is a social welfare function satisfying Arrow's Conditions
2-5 and the consistency condition for any number of alternatives, provided the preferences
of concerned individuals over every triple of alternatives is Value-Restricted, and
the number of concerned individuals for every triple is odd.}

Two concepts that are used in Theorem 1 are defined by Sen (1966) as follows.
\begin{itemize}
\item [$\diamond$] \textbf{Value of alternative:} \textit{The value of an alternative in a triple for an individual preference ordering is its characteristic of being "best," "worst," or "medium."}
\item [$\diamond$] \textbf{Value-Restricted preference pattern:} \textit{A set of individual preferences
over a triple of alternatives such that there exist one alternative and one value with
the characteristic that the alternative never has that value in any individual's preference
ordering, is called a Value-Restricted Preference pattern over that triple
for those individuals.}
\end{itemize}

Before presenting our result, we discuss the key aspects of our analysis.

An alternative in a triple being a Value-Restricted one means that this alternative is excluded from being assigned at least one position from $\{1,2,3\}$. From Definition 1 we know that an entry of a PM is a set containing an alternative's possible ranking position or positions. When the individuals are concerned ones over a triple of alternatives and their preferences over this triple are represented by PMs, if an alternative in the triple is a Value-Restricted one, then the union set of the concerned individuals' PMs' entries corresponding to this alternative will not include at least one element from $\{1,2,3\}$, i.e., in this case the cardinality number of the specified union set is less than 3.

The above analysis indicates that, Sen's condition can be stated in an alternative way. This is described explicitly next as Theorem 1'.

\textbf{Theorem 1'} \textit{Assume that the alternative set is $A=\{x_1,x_2,\ldots,x_m\}$ where $2<m<+\infty$, the individual set is
$N=\{1,2,\ldots,n\}$ where $2<n<+\infty$, and the individuals' preferences over the alternative set are weak orderings. Denote by $(x_{t_1},x_{t_2},x_{t_3})$ the triples of alternatives, where $t=1,2,\ldots,\binom{m}{3}$. Denote by $V_t$ the maximal set including concerned individuals who are  specified in terms of their preferences over that triple $(x_{t_1},x_{t_2},x_{t_3})$, and represent their preferences over that triple by PMs. Specifically, for triple $(x_{t_1},x_{t_2},x_{t_3})$ the concerned individuals' PMs are denoted by
$$
\begin{array}{c}{\hspace{0.3cm}\begin{array}{cccc}\text{Concerned individual k's PM with respect to triple } (x_{t_1},x_{t_2},x_{t_3}):
\end{array}}\\
{\begin{array}{cc}
{\begin{array}{c} x_{t_1}\\
x_{t_2}\\
x_{t_3}
\end{array}}&
{ \left[\begin{array}{c}
PM_{t_1}^{(k)}\\
PM_{t_2}^{(k)}\\
PM_{t_3}^{(k)}\end{array}\right]
}
=\left[PM_{t_i}^{(k)}\right]_{3\times 1}, k\in V_t.
\end{array}}
\end{array}
$$
The method of majority decision is a social welfare function satisfying Arrow's Conditions
2-5 and the consistency condition for any number of alternatives, provided every triple of alternatives satisfy
$$\exists t_i\left(\big|\cup_{k=1}^{|V_t|}PM_{t_i}^{(k)}\big|<3\right),|V_t| \text{ is odd}; t=1,2,\ldots,\binom{m}{3}.\eqno(3)$$
}

To illustrate, we consider an example used by Sen (1966).

\textbf{Example 2} Five voters provide the following preference orderings over four alternatives $\{w,x,y,z\}$:
$$\begin{array}{ll}
 \text{Voter 1}: & w\sim x\succ y\succ z,\\
 \text{Voter 2}: & x\sim w\succ z\succ y,\\
 \text{Voter 3}: & z\sim x\succ y\succ w,\\
 \text{Voter 4}: & z\succ y\sim x\succ w,\\
 \text{Voter 5}: & z\succ y\succ x\succ w.
\end{array}
$$
There are 4 possible alternative triples and they are all Value-Restricted, and thus the majority decision yields a transitive social preference ordering.

Here we examine the triples with Inequality (3). With regard to triple $(w,x,y)$, the voters' preferences are represented by PMs:
$$
\begin{array}{c}{\hspace{1cm}\begin{array}{cccccc}\text{Voter 1}:
&\hspace{0.2cm}\text{Voter 2}: &\hspace{0.1cm}\text{Voter 3}: &\hspace{0.02cm}\text{Voter 4}: &\hspace{0.02cm}\text{Voter 5}:
\end{array}}\\
{\begin{array}{cc}
{\begin{array}{c} w\\
x\\
y
\end{array}}&
{ \left[\begin{array}{c}
\{1,2\}\\
\{1,2\}\\
\{3\}\end{array}\right],
\left[\begin{array}{c}\{1,2\}\\
\{1,2\}\\
\{3\}\end{array}\right],
\left[\begin{array}{c}\{3\}\\
\{1\}\\
\{2\}\end{array}\right],
\left[\begin{array}{c}\{3\}\\
\{1,2\}\\
\{1,2\}\end{array}\right],
\left[\begin{array}{c}\{3\}\\
\{2\}\\
\{1\}\end{array}\right].
}
\end{array}}
\end{array}
$$

The union set of the second entries of the above PMs is $\{1,2\}$, whose cardinality number is less than 3. Hence Inequality (3) is satisfied by the triple $(w,x,y)$. Similarly, one can easily check that Inequality (3) is fulfilled by the other three triples.

\section{Sen's condition in equations}

In this section, Sen's transitivity condition will be described in such a way that the membership preference map (mPM) is employed to characterize the Value-Restricted requirement.

In Theorem 1', the concerned individual's preferences over a triple of alternatives are represented by PMs. From Definition 2 we know that a PM has a one-to-one correspondence of mPM. A very natural way is to further represent the PMs in Theorem 1' by mPMs and the thoerem will then transform into another form. Actually, since an element of a mPM indicates whether an alternative can be ranked at a certain position or not, thus if a certain alternative in a triple is Value-Restricted, then none of the concerned individuals would like to rank the certain alternative at a certain position, i.e., in this case the 0-1 element that corresponds to the certain alternative and the certain ranking position of each concerned individual's mPM is 0. As a consequence of this analysis, the problem of examining inequalities for a transitive social ordering is equivalent to checking a set of equations. This is described explicitly next as Theorem 1".

\textbf{Theorem 1"} \textit{Assume that the alternative set is $A=\{x_1,x_2,\ldots,x_m\}$ where $2<m<+\infty$, the individual set is
$N=\{1,2,\ldots,n\}$ where $2<n<+\infty$, and the individuals' preferences over the alternative set are weak orderings. Denote by $(x_{t_1},x_{t_2},x_{t_3})$ the triples of alternatives, where $t=1,2,\ldots,\binom{m}{3}$. Denote by $V_t$ the maximal set including concerned individuals who are  specified in terms of their preferences over that triple $(x_{t_1},x_{t_2},x_{t_3})$, and represent their preferences over that triple by PMs. Specifically, for triple $(x_{t_1},x_{t_2},x_{t_3})$ the concerned individuals' PMs are denoted by
$$
\begin{array}{c}{\hspace{0.3cm}\begin{array}{cccc}\text{Concerned individual k's PM with respect to triple } (x_{t_1},x_{t_2},x_{t_3}):
\end{array}}\\
{\begin{array}{cc}
{\begin{array}{c} x_{t_1}\\
x_{t_2}\\
x_{t_3}
\end{array}}&
{ \left[\begin{array}{c}
PM_{t_1}^{(k)}\\
PM_{t_2}^{(k)}\\
PM_{t_3}^{(k)}\end{array}\right]
}
=\left[PM_{t_i}^{(k)}\right]_{3\times 1}, k\in V_t.
\end{array}}
\end{array}
$$
The concerned individual k's PM is further represented by a mPM:
$$
\begin{array}{c}{\hspace{0.3cm}\begin{array}{cc}\text{Concerned individual k's mPM with respect to triple } (x_{t_1},x_{t_2},x_{t_3}):
\end{array}}\\
\begin{array}{c}{\hspace{-0.7cm}\begin{array}{cc}\{1\}\hspace{1.3cm}\{2\}\hspace{1.3cm}\{3\}
\end{array}}\\
{\begin{array}{cc}
{\begin{array}{c} x_{t_1}\\
x_{t_2}\\
x_{t_3}
\end{array}}&
{ \left[\begin{array}{ccc}
mPM_{t_{1,1}}^{(k)} &  mPM_{t_{1,2}}^{(k)} &  mPM_{t_{1,3}}^{(k)} \\
mPM_{t_{2,1}}^{(k)} &  mPM_{t_{2,2}}^{(k)} &  mPM_{t_{2,3}}^{(k)} \\
mPM_{t_{3,1}}^{(k)} &  mPM_{t_{3,2}}^{(k)} &  mPM_{t_{3,3}}^{(k)}
\end{array}\right]_{3\times 3},
}
\end{array}k\in V_t.}
\end{array}
\end{array}
$$
The method of majority decision is a social welfare function satisfying Arrow's Conditions
2-5 and the consistency condition for any number of alternatives, provided every triple of alternatives satisfy
$$\min_i\min_j\left\{\sum_{k=1}^{|V_t|}mPM_{t_{i,j}}^{(k)}\right\}=0,|V_t| \text{ is odd}; t=1,2,\ldots,\binom{m}{3}.\eqno(4)$$
}

To illustrate, we reconsider Example 2.

With regard to triple $(w,x,y)$, the voters' mPMs are:
$$
\begin{array}{c}{\hspace{0.6cm}\begin{array}{cccccc}\text{Voter 1}:
&\hspace{0.6cm}\text{Voter 2}: &\hspace{0.5cm}\text{Voter 3}: &\hspace{0.5cm}\text{Voter 4}: &\hspace{0.5cm}\text{Voter 5}:
\end{array}}\\
\begin{array}{c}{\hspace{0.8cm}\begin{array}{cccccc}\{1\}\hspace{0.08cm}\{2\}\hspace{0.08cm}\{3\}
&\hspace{0.1cm}\{1\}\hspace{0.08cm}\{2\}\hspace{0.08cm}\{3\} &\hspace{0.1cm}\{1\}\hspace{0.08cm}\{2\}\hspace{0.08cm}\{3\}
&\hspace{0.1cm}\{1\}\hspace{0.08cm}\{2\}\hspace{0.08cm}\{3\}
&\hspace{0.1cm}\{1\}\hspace{0.08cm}\{2\}\hspace{0.08cm}\{3\}
\end{array}}\\
{\begin{array}{cc}
{\begin{array}{c} w\\
x\\
y
\end{array}}&
{ \left[\begin{array}{ccc}
1 &  1 &  0 \\
1 &  1 &  0 \\
0 &  0 &  1
\end{array}\right],
\left[\begin{array}{ccc}
1 &  1 &  0 \\
1 &  1 &  0 \\
0 &  0 &  1
\end{array}\right],
\left[\begin{array}{ccc}
0 &  0 &  1 \\
1 &  0 &  0 \\
0 &  1 &  0
\end{array}\right],
\left[\begin{array}{ccc}
0 &  0 &  1 \\
1 &  1 &  0 \\
1 &  1 &  0
\end{array}\right],
\left[\begin{array}{ccc}
0 &  0 &  1 \\
0 &  1 &  0 \\
1 &  0 &  0
\end{array}\right].
}
\end{array}}
\end{array}
\end{array}
$$
The sum of the above five 0-1 matrices is:
$$\left[\begin{array}{ccc}
2 &  2 &  3 \\
4 &  4 &  0 \\
2 &  2 &  2
\end{array}\right],$$
from which one can see that Eq.(4) is fulfilled in that the $(2,3)th$ element of the sum matrix is 0. Likewise, it is easy to check that Eq.(4) is fulfilled by the other three triples.

\section{Concluding remarks}

Sen's transitivity condition is a sort of qualitative description. This article attempted to describe it in a quantitative manner. Our way to do this was to represent the concerned individual's preferences over a triple of alternatives by preference maps (PMs) and membership preference maps (mPMs). Each of these presented a theorem establishing a quantitative description of Sen's value restriction condition constrained on concerned individuals' preferences for the majority decision to yield a transitive social preference ordering. The practical significance of our results lies in the testing of Sen's transitivity condition with a computer. This is particularly true when, as presented in Theorem 1", the condition is constructed based on 0-1 matrices.

\vspace{0.3cm}
% \textbf{Acknowledgments}

%The work was supported by the National Natural Science Foundation of China (No. ).

\end{document}